# Disorder-Induced Suppression of Antiferromagnetic Magnon Transport in $Cr_2O_3$

Josiah Keagy[1], Haoyu Liu[1], Junyu Tang[1], Weilun Tan[1], Wei Yuan[1], Sumukh Mahesh[1], Ran Cheng[1,2], and Jing Shi[1]

1. Department of Physics & Astronomy, University of California, Riverside, CA 92521, USA
2. Department of Electrical & Computer Engineering, University of California, Riverside CA 92521, USA

We explore the impact of spin disorder associated with structural defects on antiferromagnetic magnon transport by probing the spin-flop transition of $Cr_2O_3$ using spin Seebeck effect measurements. By fabricating homoepitaxial $Cr_2O_3$ films grown on smooth $Cr_2O_3$ crystals, we systematically vary the thickness of the films in which the presence of point defects modulates the spin-flop transition field. We find that magnon propagation through the film is entirely suppressed for film thickness greater than 3 nm, revealing the pivotal role of disorder in governing antiferromagnetic magnon transport.

Antiferromagnetic magnon transport has garnered significant interest due to its potential applications in spintronics and spin-based quantum information technologies [1–9]. The spin Seebeck effect [10] (SSE) is a well-established technique for studying thermally driven magnon transport in magnetic systems [11–15]. In uniaxial antiferromagnetic (AFM) insulators such as $Cr_2O_3$, $MnF_2$, and $FeF_2$, the diffusion of circularly polarized magnons under an applied magnetic field generates a net angular momentum flow, producing characteristic temperature and magnetic field dependencies in the SSE signal [15–17]. Within the standard SSE framework [17,18], a thermal gradient in the AFM source material generates a magnon spin current, which is converted into an electric voltage by an adjacent heavy metal (HM) layer, such as Pt, via the inverse spin Hall effect. This voltage signal is typically attributed to intrinsic material properties [19,20], including spin polarization arising from an imbalance of left-handed (LH) and right-handed (RH) magnons [21–23], spin-mixing conductance [24–26], and the spin Hall angle [27] of the detector. Despite the well-established role of intrinsic factors, the influence of extrinsic factors, such as material defects-induced spin disorder, on magnon transport and the resulting SSE signal remains largely unexplored. While nonlocal measurements have demonstrated long-range AFM magnon transport via mechanisms such as diffusion or spin superfluid [28–30], these studies were conducted in systems with fixed disorder configurations. This limitation has precluded systematic investigations into the influence of extrinsic effects, highlighting the need for controlled studies to disentangle intrinsic and extrinsic contributions to SSE signals.

In this study, we address this gap by systematically probing the extrinsic contributions to the SSE in $Cr_2O_3$. We first separate extrinsic effects from intrinsic properties in bulk $Cr_2O_3$ crystals. Subsequently, we introduce homoepitaxial $Cr_2O_3$ films of varying thickness, each exhibiting distinct SSE characteristics due to structural defects. By comparing the SSE responses, we directly identify the contributions of bulk magnons and study the effect of extrinsic disorder on magnon transport.

We perform SSE measurements on $Cr_2O_3$ crystals with ($10\bar{1}0$) and ($11\bar{2}0$) orientations, where the c-axis lies within the sample plane. The experiments utilize the standard longitudinal SSE geometry (inset of Figure 1a). As the magnetic field $H$ is swept along the c-axis, the SSE signal magnitude is expected to increase monotonically because of the increased LH magnon populations as the field strength increases. At the spin-flop transition (SFT) at 6 T, the LH magnon energy reaches zero. Above the SFT, a RH circular eigenmode emerges, which gives rise to an abrupt SSE signal reversal due to the switch in magnon chirality from LH to RH magnons [17,20]. Simulations were conducted, adopting an SSE model consistent with that presented in Ref. 20, which accounted for all AFM magnons by assuming a linear dispersion relation across the Brillouin zone. These simulations were performed in an idealized context, excluding extrinsic factors such as interfacial defects. At low temperatures, calculations indicate a significant reduction in SSE signal magnitude above the SFT (lower panel of Figure 1a). As anticipated, the measured SSE signal exhibits the characteristic jump at the SFT in both crystals (Figure 1a); however, contrary to theoretical expectations, a substantial signal persists above the SFT field. Notably, this above-SFT signal magnitude varies significantly across samples, suggesting an extrinsic origin. This interpretation aligns with earlier reports [14,17], where SSE responses in $Cr_2O_3$ samples with ion bombarded surfaces deviated markedly from those in the smooth, pristine samples.

To further investigate the origin of the above-SFT SSE signal, we examine the SSE behavior in the hard-axis configuration, where the in-plane magnetic field is applied perpendicular to the c-axis (Figure 1b). In this setup, the field induces a canted magnetic moment ***m***, and the magnon energy gap increases monotonically with the field strength, similar to the behavior above the SFT in the easy-axis configuration. As demonstrated by the calculations for both configurations in Figure S1 (Supplemental Material [31]), the intrinsic magnon contribution in the hard-axis configuration is negligibly small. Thus, any observed SSE signal in this configuration likely arises from extrinsic effects. As shown in Figure 1b, fabricated devices feature two orthogonal Pt detection channels: one parallel and the other perpendicular to the c-axis. The Pt films are deposited by sputtering onto $Cr_2O_3$ crystals prepared to have different surface conditions. Each Pt channel measures either 2.9 mm or 1.9 mm in length, 180 μm or 15 μm in width, and 3.5 nm in thickness, with dimensions precisely defined by photolithography and Ar-plasma etching. We measure both the easy-axis and hard-axis magnetic field responses by recording SSE voltages in the Pt channel that is perpendicular to the applied field. To drive the SSE signal in each measurement, a 15 nm-thick Pt heater is employed (not shown), vertically aligned with the selected Pt detection channel, similar to single-channel measurements. The heater is separated from the Pt detection layer by a ~100 nm thick insulating $Al_2O_3$ layer deposited by atomic layer deposition, which produces controlled temperature gradient. To introduce surface conditions with varying extrinsic defect levels, we implement an iterative process. Subsequent to the completion of the initial SSE measurements, the devices are removed through mechanical polishing with 6 μm-sized diamond suspension. The exposed $Cr_2O_3$ surface is then subjected to chemical-mechanical polishing [32] to generate a fresh, smooth surface. Repeated chemical mechanical polishing is performed to achieve desired surface smoothness. New SSE devices with identical geometry are then fabricated on the polished $Cr_2O_3$ surface. This process is repeated multiple times, with the root-mean-square (RMS) roughness of each $Cr_2O_3$ surface characterized using atomic force microscopy.

Figures 1c presents the 10 K easy-axis (upper) and hard-axis (lower) SSE signals for three representative surface conditions of a $(11\bar{2}0)$ bulk crystal with RMS roughness values of 0.50 nm, 0.26 nm, and 0.20 nm. Several key contrasts emerge. First, in the easy-axis configuration, the SSE response for the smoothest surface closely resembles the commonly observed SSE responses in bulk crystals with smooth surfaces with comparable RMS roughness. While abrupt SFT jumps are evident in all samples, the slope of the field dependence below the SFT changes dramatically, ranging from positive for the smoothest surface to negative for the roughest. Additionally, the above-SFT SSE signal increases with increasing RMS roughness, underscoring the impact of extrinsic factors on SSE behavior. Second, the hard-axis SSE signals exhibit a comparable magnitude to the above-SFT easy-axis SSE signals, suggesting a shared origin unrelated to intrinsic magnon contributions (Figure S1, Supplemental Material [31]). Similar to the above-SFT easy-axis signals, the hard-axis SSE signal magnitude increases with RMS roughness. Third, the hard-axis SSE responses can be interpreted as paramagnetic in nature. Notably, the hard-axis responses are well-fitted using the Brillouin function [33,34] with weak antiferromagnetic interactions (Figure S2, Supplemental Material [31]), indicative of weakly interacting spins near the $Cr_2O_3$ surface. In perfect crystal, the $(11\bar{2}0)$ surface is magnetically compensated; however, rough surfaces may introduce uncompensated magnetic moments behaving as nearly paramagnetic spins. Due to their isotropic characteristics, these paramagnetic signals overlay the intrinsic easy-axis responses, distorting them when the background signal is substantial. To correct this effect, we subtract the hard-axis contributions from the easy-axis response. The corrected intrinsic SSE signals, $\Delta\bar{V}_{SSE}$, shown in Figure 1d, align more closely with the expected AFM magnon SSE responses [20].

After removing the primary background, secondary features become apparent among the three samples. Figure 1d reveals a clear trend: the data for the samples with 0.26 nm and 0.20 nm RMS roughness exhibit comparable signal magnitudes and sharp SFT transitions at 6 T. In contrast, the 0.50 nm curve, representing the roughest surface, shows a significantly reduced SSE signal and a noticeably broader SFT feature. These results suggest that the $Cr_2O_3$ surface harbors unique properties not fully captured by RMS roughness alone. The absence of a sharp SFT feature in the roughest sample indicates that bulk magnons are unable to efficiently reach the $Cr_2O_3$/Pt interface, likely due to enhanced scattering by defects embedded in the region beneath the $Cr_2O_3$ surface.

To further investigate the impact of extrinsic factors beyond paramagnetic spins on magnon transport, we adopt a controlled approach by growing thin epitaxial $Cr_2O_3$ layers using pulsed laser deposition (PLD) on atomically flat substrates to ensure smooth surfaces. This method enables precise control over film thickness, facilitating the study of evolving structural and magnetic properties. We employ a $Cr_2O_3$ target in an oxygen environment to grow the $Cr_2O_3$ films, systematically varying parameters such as pressure (0.5-2 mTorr of $O_2$), substrate temperature (500-530 °C), laser energy (150-200 mJ) and repetition rate (1-2 Hz), and post-deposition annealing conditions (with and without) (results from representative samples are shown in Figure S3, Supplemental Material [31]). Figure 2a shows the atomic force microscopy image of a typical film grown epitaxially on an $Al_2O_3(11\bar{2}0)$ substrate, hereafter referred as a heteroepitaxial film. The image reveals clear atomic terraces and an RMS roughness of 0.15 nm. Both the linecut across atomic terraces (above) and a streaky reflection high-energy electron diffraction (RHEED) pattern (inset) confirm high-quality growth. The overall epitaxial quality is validated by the appearance of Kiessig fringes around the $(11\bar{2}0)$ Bragg peak in the X-ray diffraction pattern (Figure 2b), and the presence of dozens of oscillations in the X-ray reflectivity spectrum (inset of Figure 2b), both indicative of smooth film surfaces and strong structural coherence. Furthermore, high-resolution transmission electron microscopy (HRTEM) cross-sectional view (Figure 2c) reveals well-extended atomic layers throughout the film, consistent with bulk-like crystallinity. Figure 2d presents the SSE data for two representative heteroepitaxial films after the paramagnetic-like background is removed (detailed information in Figure S4, Supplemental Material [31]). While the overall characteristics resemble those of the bulk crystals, the sharp SFT feature observed in bulk crystals with smooth surfaces is absent. Instead, the much-broadened SSE responses mirror that seen in bulk crystal with the roughest surface (RMS ~ 0.50 nm). Given the atomically flat surface of the heteroepitaxial films, this broadening suggests the presence of microscopic imperfections inside the films affecting the magnetic properties, akin to defects located beneath the rough surface of the bulk sample. The HRTEM image in Figure 2c shows faint variations in background intensity over a ~ 10 nm length scale, indicative of microscopic structural disorder despite the overall epitaxial nature of the films. We hypothesize that this disorder induces variability in magnetic anisotropy or exchange strength, leading to a broader distribution of SFT fields. We find that these characteristics appear stable and relatively insensitive to adjustments in growth parameters. This suggests a robust defect configuration, consistent with findings reported by others [35].

The results from the sample with a very rough crystal surface and PLD-grown heteroepitaxial films point to a common origin for the broad SFT distribution. A key question that arises is how deep the $Cr_2O_3$ surface layer extends and how it affects transport of magnons from the underlying bulk, which is relatively disorder-free. To address this question, we employ homoepitaxy with PLD to grow $Cr_2O_3$ layers with varying thicknesses on smooth (< 0.3 nm RMS roughness), polished $Cr_2O_3$ crystals. To ensure a high degree of comparability, we divide the 5 mm x 5 mm crystal surface into two sections. One section is covered with a homoepitaxial film of a specific thickness, while the other section is left pristine to serve as a bulk reference. Both sections undergo identical SSE device fabrication processes. For the thickness dependence study, we fabricate eight SSE devices. Four of these devices incorporate homoepitaxial films with distinct thicknesses (3, 9, 14, and 26 nm), while the remaining four serve as their respective bulk references. The PLD-grown films exhibit excellent structural properties as evidenced by atomic force microscopy images (Figure S5, Supplemental Material [31]), comparable with or better than those of the heteroepitaxial films. In all bulk reference devices, we observe minor variations in the SSE signal magnitude; however, the overall SSE response shape remains consistent, and especially, the sharp SFT jump occurs at 6 T as shown in Figure 1. On the same polished $Cr_2O_3$ surface, both the bare bulk surface and homoepitaxial films show a high uniformity as shown in Figure S6. To correct the signal magnitude variations across different polished samples, we scale the SSE signal from each homoepitaxial film device by the maximum signal from its local reference device.

Figure 3a presents the scaled SSE data for four homoepitaxial $Cr_2O_3$ film devices of varying thicknesses and a representative bulk reference device (companion to the 3 nm film device). Clearly, the 3 nm device exhibits an SFT field nearly identical to that of the bulk with a slightly smaller signal magnitude, indicating minimal attenuation of bulk magnons through the ultrathin film. This observation is consistent with the high structural quality of the 3 nm film, confirmed by the absence of pinholes or cracks imaged by atomic force microscopy. As the epitaxial layer thickness increases to 9 nm, however, the SFT field shifts to ~6.7 T, representing a ~12% increase relative to the bulk. Additionally, the characteristic SSE jump at the bulk SFT field (6 T) disappears. These results suggest that bulk magnons no longer contribute to the detected SSE signal in the 9 nm film. Instead, magnons originating within the $Cr_2O_3$ layer itself dominate the SSE response. This shift in magnon transport behavior reflects a crossover where the $Cr_2O_3$ surface layer effectively blocks bulk magnon propagation due to increased scattering or attenuation.

Further changes are observed with increasing film thickness as illustrated in Figure 3c. For the 26 nm device, the SFT distribution broadens so extensively that the transition remains incomplete even at 14 T. This broadening coincides with a marked reduction in SSE magnitude. Together, these observations indicate that, in films thicker than 3 nm, bulk magnons are unable to propagate through the shallow $Cr_2O_3$ surface layer to reach the Pt detector. Consequently, the measured SSE signal arises exclusively from the film itself. Figure 3b provides further evidence by comparing homoepitaxial and heteroepitaxial $Cr_2O_3$ films of comparable thickness (12 and 14 nm, respectively). Despite the presence of bulk magnons underneath the homoepitaxial films, their SSE characteristics - broadened and upshifted SFT - are indistinguishable from those of the heteroepitaxial films. This strongly suggests that bulk magnons are effectively scattered or absorbed by the thin $Cr_2O_3$ surface layer and the detected SSE signals are dominated by the $Cr_2O_3$ films. The presence of microscopic structural defects in both film types likely accounts for this behavior. These defects perturb the local magnetic anisotropy and/or exchange interactions [36,37], leading to spatially varying SFT fields. In homoepitaxial films, such scattering of bulk magnons by defects renders the underlying $Cr_2O_3$ crystal effectively indistinguishable from a non-magnetic substrate. As a result, the SSE response originates solely from the disordered $Cr_2O_3$ film, which exhibits a broad SFT distribution.

To understand the origin of this strikingly short magnon diffusion length, we now turn to the role of structural defects. In heteroepitaxial $Cr_2O_3$ films grown on $Al_2O_3$ substrates, lattice mismatch (~4% compressive strain along the c-axis) can induce defects such as dislocations and misorientations, which distort the regular lattice structure. Such defects may explain the contrast variations observed in Figure 2c. Homoepitaxial films, in principle, should yield superior quality due to the absence of lattice mismatch, as seen in other systems like Si/Si [38]. However, Figure 4a shows that even in homoepitaxial films, while crystal planes appear continuous across the interface in the HRTEM image, underlying contrast variations - similar to those in heteroepitaxial films - are still evident. Other defects such as dislocations are very rare. This suggests that embedded point defects, not resolved by HRTEM, contribute significantly to these variations. Oxygen deficiency is a known type of point defect in oxide films [39,40]. Figure 4b shows the thickness profile of oxygen deficiency, measured via energy-dispersive X-ray (EDX) spectroscopy in a transmission electron microscope where the HRTEM images were acquired, alongside SFT fields and SSE amplitude in devices of varying thickness. While the Cr/O ratio starts approximately at the expected value of 2/3 at the interface, it gradually increases as the thickness increases, rising by over 20% at a depth of 26 nm. Missing oxygen atoms cause lattice distortions and disrupt the super-exchange interaction between AFM spins, weakening the AFM exchange. Surprisingly, this does not explain increased SFT fields in thicker films. Instead, oxygen vacancies may induce contraction along the c-axis, which, through the magnetoelastic effect [41], enhances the magnetic anisotropy and shifts the SFT to higher fields. Additionally, these structural defects can scatter magnons via magnetoelastic interaction, attenuating the SSE magnitude. To fully elucidate the underlying mechanisms, further theoretical calculations are required.

In summary, this systematic investigation identifies two key extrinsic factors affecting the intrinsic SSE responses from bulk magnons: paramagnetic-like SSE behavior and oxygen deficiency-induced point defects. By systematically varying the thickness of homoepitaxial films, we have demonstrated that these defects are highly effective in suppressing magnon spin current transport, thereby reducing the magnon diffusion length to just a few nanometers. This highlights the critical role of structural and chemical defects in governing magnon dynamics and underscores the need for precise control over material quality for AFM spintronics and AFM magnonics applications which require long-distance magnon transport for angular momentum and quantum information propagation.

We thank Igor Barsukov, Benedetta Flebus, Jayakrishnan Muttathil Prabhakarapada Nair, for useful discussions. This work is supported in part by NSF-DMR-2203134.

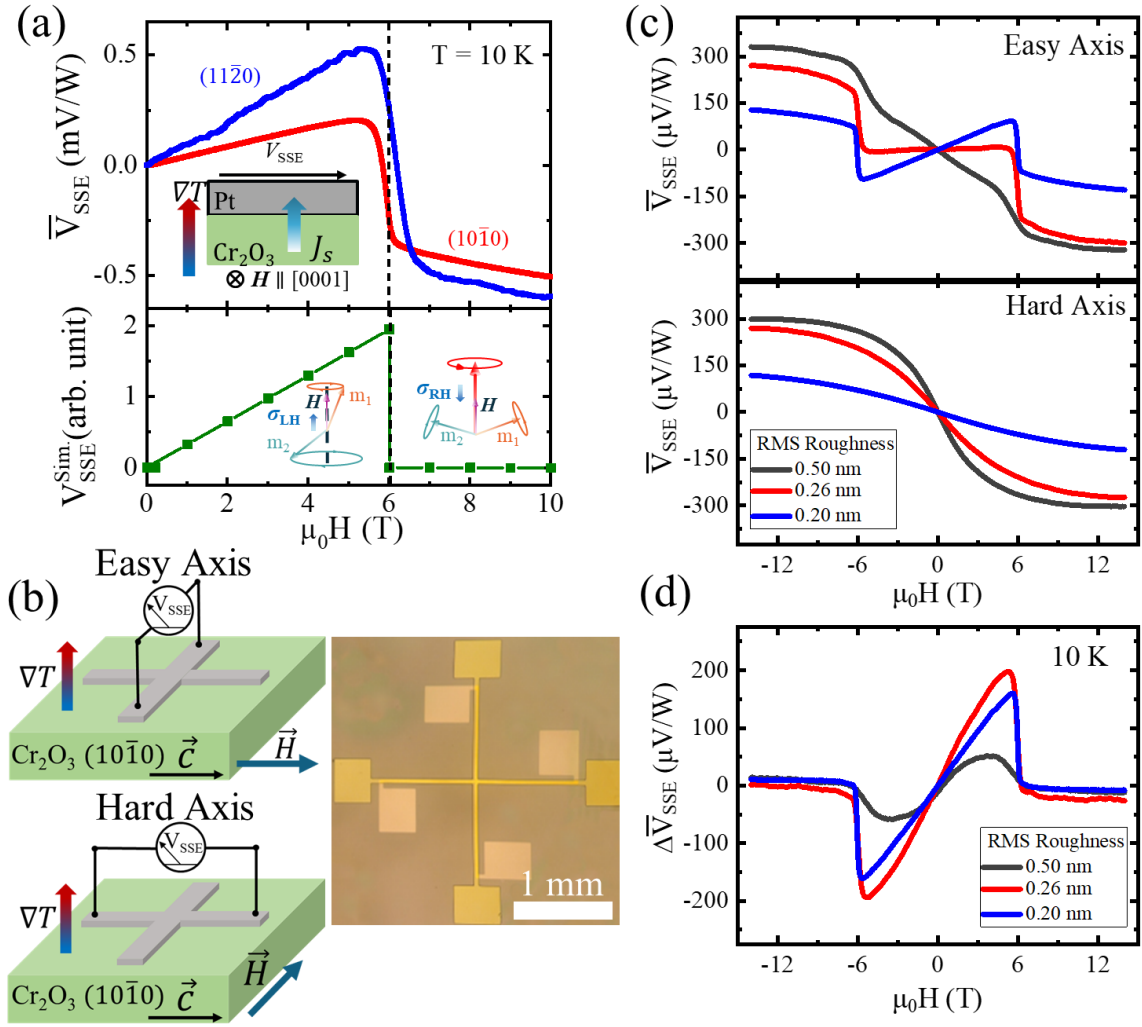

**Figure 1. Influence of Surface Preparation on SSE response in $Cr_2O_3$ Bulk Crystals** (a) Inset: Schematic of the longitudinal SSE measurement setup, where a spin current ($J_S$) induced by a vertical temperature gradient ($\nabla T$) is converted into an electric voltage ($V_{SSE}$) by a HM such as Pt. Top graph: SSE responses ($V_{SSE}/P$, $P$: heater power) measured at 10 K for two bulk $Cr_2O_3$ crystals with $(10\bar{1}0)$ and $(11\bar{2}0)$ orientations, both with compensated surfaces. The RMS roughness for both crystals is not greater than 0.2 nm. Bottom graph: Corresponding calculated SSE results. $\sigma_{LH}$ and $\sigma_{RH}$ are spin polarizations from LH and RH magnon, respectively, and $m_1$ and $m_2$ are sub-lattice magnetic moments. More details in Figure S1, Supplemental Material [31]. The sharp SFT at 6 T marks the boundary between LH-dominated and RH magnon SSE contributions. (b) Schematic representation of the two measurement geometries (left) and optical image of the SSE device (right). (c) SSE field sweep data (10 K) along the easy (above) and hard (below) axes for $(11\bar{2}0)$ crystal prepared under three conditions, as indicated by their RMS roughness values. (d) Difference curves ($\Delta \bar{V}_{SSE} = \Delta(\frac{V_{SSE}}{P})$) between easy-axis and hard-axis measurements shown in (c).

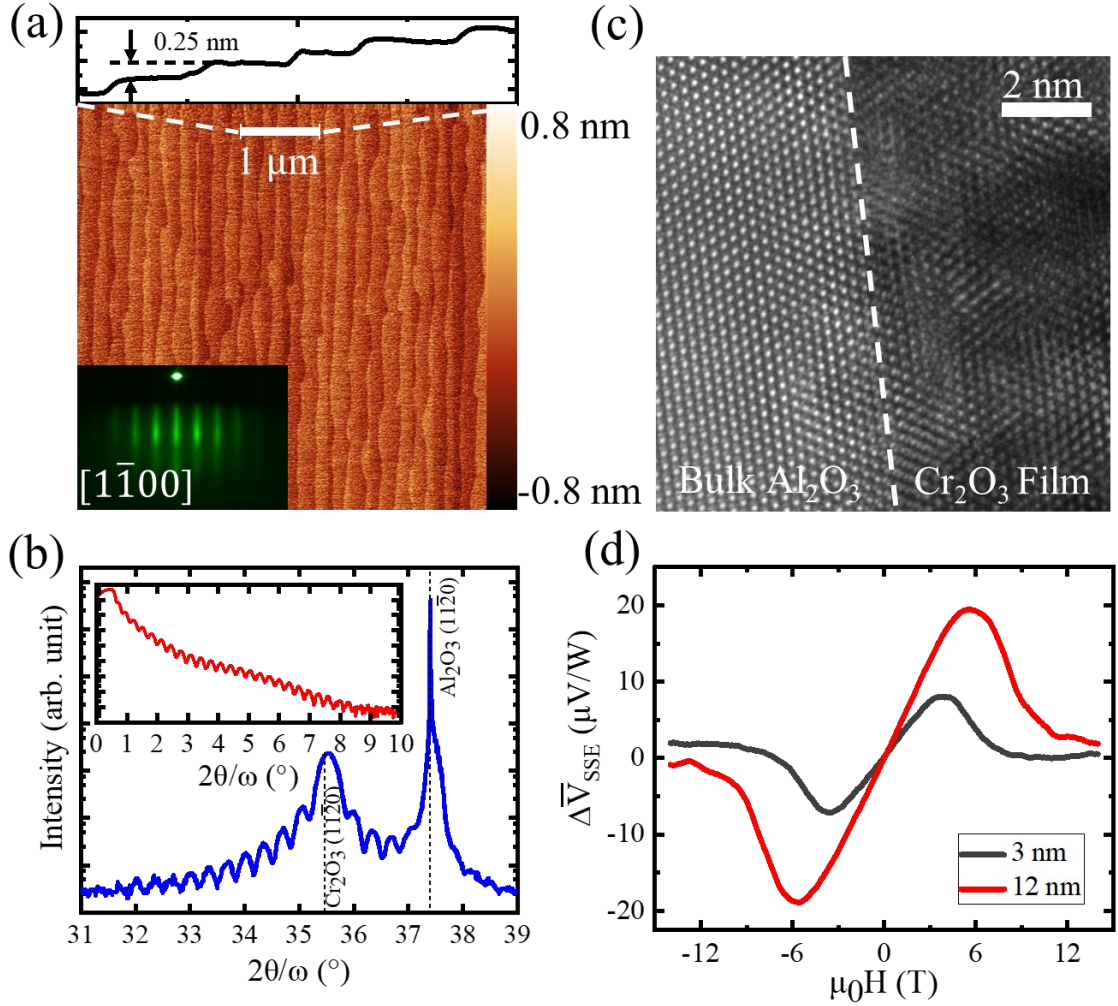

**Figure 2. Structural Characterization of $Cr_2O_3$ Films and Corresponding SSE Responses.** (a) Atomic force microscopy image of a representative $Cr_2O_3$ film grown on $Al_2O_3(11\bar{2}0)$ using PLD, with the linecut (above) illustrating the profile of atomic terraces over 1 μm range. The data is flattened to enhance the visibility of step features. The inset is a RHEED pattern obtained post-growth. (b) X-ray diffraction $\frac{2\theta}{\omega}$ scan displaying Bragg peaks corresponding to $Cr_2O_3(11\bar{2}0)$ and $Al_2O_3(11\bar{2}0)$. The presence of multiple Kiessig fringes confirms high structural coherence and flat surfaces. The inset features X-ray reflectivity data, indicating a highly uniform film. (c) Cross-sectional HRTEM image of $Al_2O_3/Cr_2O_3$ interface, demonstrating excellent epitaxy. Regions with contrast variations are visible, suggesting localized structural differences. (d) SSE difference data $\Delta\bar{V}_{SSE}$ measured at 10 K for two $Cr_2O_3$ films grown on $Al_2O_3(11\bar{2}0)$

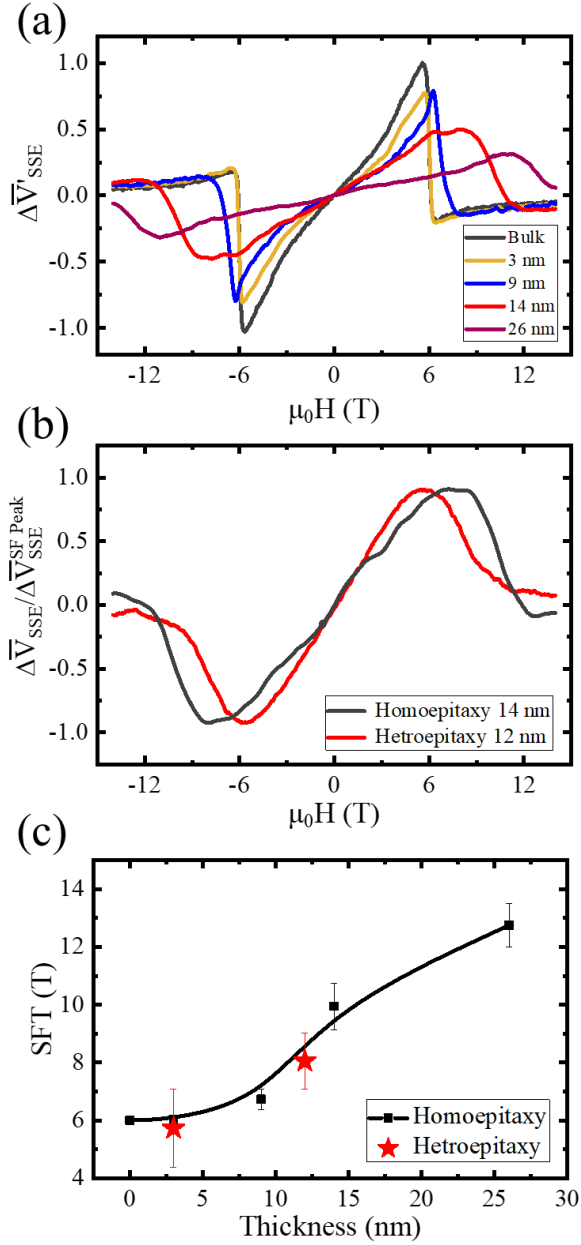

**Figure 3. SSE Data for Homoepitaxial $Cr_2O_3$ Films.** (a) $\Delta \bar{V}_{SSE}$ data of four homoepitaxial thin films of varying thickness, scaled by the peak SSE value at the SFT of the companion bulk $Cr_2O_3$ crystal device, measured at 1.8 K. Each curve is the average of up and down field sweeps. Data also includes the SSE curve from a bulk $Cr_2O_3$ crystal without an overlaying film. (b) Comparison of $\Delta \bar{V}_{SSE}$ between homoepitaxial and heteroepitaxial films at 10 K, highlighting similarities in response. Signals are normalized by their peak values. (c) Mean SFT field as a function of film thickness, measured at 10 K. Data points for two heteroepitaxial films are represented by red stars.

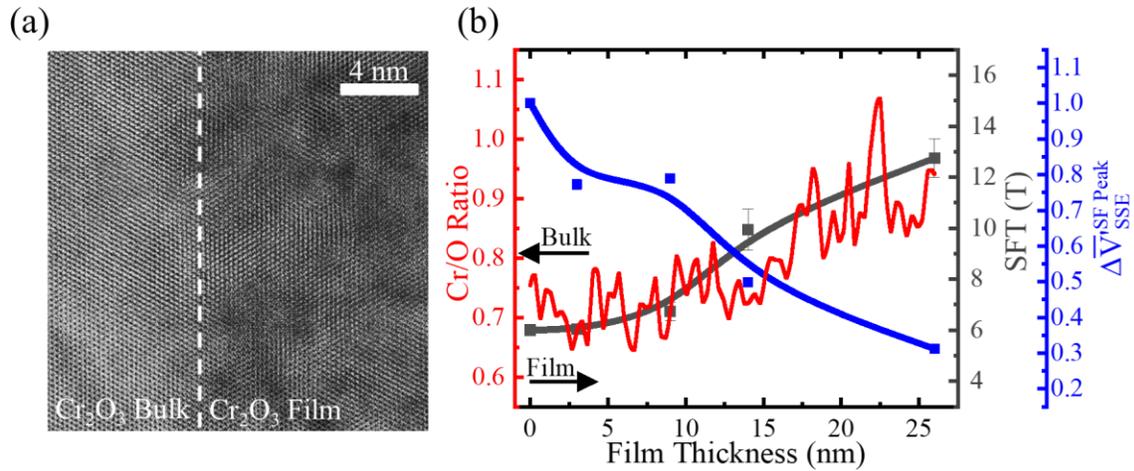

**Figure 4. Structural Defects and Oxygen Deficiency in Homoepitaxial Films.** (a) HRTEM image of a 26 nm homoepitaxial $Cr_2O_3$ film, illustrating structural continuity with visible contrast variations likely caused by defects. (b) Cr/O ratio in a homoepitaxial $Cr_2O_3$ film as a function of distance from the bulk $Cr_2O_3$ crystal surface, measured by EDX. Overlaid are the corresponding SFT fields and normalized SSE peak magnitudes plotted against homoepitaxial film thickness, highlighting the interplay between oxygen deficiency and SSE response.